\documentclass[journal,10pt]{IEEEtran}
\usepackage{subfigure}
\PassOptionsToPackage{subfigure}{tocloft}
\usepackage{CJK}
\usepackage{algorithm}
\usepackage{algorithmic}
\usepackage{amsmath}
\usepackage{amssymb}
\usepackage{graphicx}
\usepackage{epstopdf}
\usepackage{multirow}
\usepackage{stfloats}
\usepackage{cite}
\usepackage{color}
\usepackage[normalem]{ulem}
\usepackage{enumerate}
\usepackage{bbm}

\usepackage{times,amsmath,amsthm,color,amssymb,graphicx,epsfig,algorithm,algorithmic,bm,multirow,cite}
\newcommand{\correct}[2]{{\color{red}{\sout{#1}}}{\color{black}{#2}}} 
\theoremstyle{definition}

\newtheorem{proposition}{Proposition}

\newtheorem{remark}{Remark}

\floatname{algorithm}{Algorithm}

\title{Active Reconfigurable Intelligent Surface Aided Wireless Communications}

\author{Ruizhe Long, Ying-Chang Liang, \IEEEmembership{Fellow, IEEE}, Yiyang Pei, and Erik G. Larsson, \IEEEmembership{Fellow, IEEE}
\thanks{

This work has been submitted to the IEEE for possible publication. Copyright may be transferred without notice, after which this version may no longer be accessible.

This work is supported by National Natural Science Foundation of China under Grants 61631005 and U1801261, by National Key Research and Development Program of China under Grant 2018YFB1801105, and by 111 International Collaboration Project under Grant B20064.

R.~Long is with the National Key Laboratory of Science and Technology on Communications, and with the Center for Intelligent Networking and Communications (CINC), University of Electronic Science and Technology of China (UESTC), Chengdu 611731, China (e-mail: ruizhelong@gmail.com).

Y.-C. Liang is with the Center for Intelligent Networking and Communications (CINC), University of Electronic Science and Technology of China (UESTC), Chengdu 611731, China (e-mail: liangyc@ieee.org).

Y. Pei is with the Singapore Institute of Technology, 10 Dover Drive, Singapore 138683, Singapore (e-mail: Yiyang.Pei@singaporetech.edu.sg).

E. G. Larsson is with the Department of Electrical Engineering (ISY), Link$\ddot{\rm o}$ping University, SE-581 83 Link$\ddot{\rm o}$ping, Sweden (email: erik.g.larsson@liu.se).
}}

\begin{document}

\maketitle
\begin{abstract}
Reconfigurable Intelligent Surface (RIS) is a promising solution to reconfigure the wireless environment in a controllable way. To compensate for the double-fading attenuation in the RIS-aided link, a large number of passive reflecting elements (REs) are conventionally deployed at the RIS, resulting in large surface size and considerable circuit power consumption. In this paper, we propose a new type of RIS, called active RIS, where each RE is assisted by active loads (negative resistance), that reflect and amplify the incident signal instead of only reflecting it with the adjustable phase shift as in the case of a passive RIS. Therefore, for a given power budget at the RIS, a strengthened RIS-aided link can be achieved by increasing the number of active REs as well as amplifying the incident signal. We consider the use of an active RIS to a single input multiple output (SIMO) system. {However, it would unintentionally amplify the RIS-correlated noise, and thus the proposed system has to balance the conflict between the received signal power maximization and the RIS-correlated noise minimization at the receiver. To achieve this goal, it has to optimize the reflecting coefficient matrix at the RIS and the receive beamforming at the receiver.} An alternating optimization algorithm is proposed to solve the problem. Specifically, the receive beamforming is obtained with a closed-form solution based on linear minimum-mean-square-error (MMSE) criterion, while the reflecting coefficient matrix is obtained by solving a series of sequential convex approximation (SCA) problems. Simulation results show that the proposed active RIS-aided system could achieve better performance over the conventional passive RIS-aided system with the same power budget.
\end{abstract}

\section{Introduction}
\label{sec:intro}
Reconfigurable intelligent surface (RIS), driven by the reflective radio, has recently attracted great attentions from both academia and industry, due to its potential to intelligently reconfigure wireless communication environment with cost-effective reflecting elements (REs) \cite{Liang2019,Liang2020Symbiotic,ElMossallamy2020Reconfigurable}. Specifically, based on the electromagnetic (EM) scattering principles, each RE is able to reflect the incident radio frequency (RF) signal with an adjustable reflecting coefficient (amplitude and phase). With the proper design of the reflecting coefficients at each RE, RIS is able to enhance the signal reception at the desired destinations and/or suppress the interference to unintended users \cite{Renzo2020,zhang2020large}, thereby artificially creating a favorable propagation condition and enhancing the existing wireless communication without requiring any bulkier RF chain \cite{DiRenzo2019}. Thus, compared to the traditional wireless communication with complex RF chain components, RIS has emerged as one of the key enabler to achieve more spectrum- and energy-efficient wireless communications.

The applications of RIS in wireless communication have recently sparkled a flurry of research activities. RIS usually serves as a reflector to assist the existing communication by reforming the reflected signal in a customized manner \cite{Huang2019,Guo2020,Wu2019,Chen2019,yuan2019intelligent,xu2020resource}. With the presence of RIS, wireless communication systems can achieve great improvements in the energy and spectrum efficiency \cite{Huang2019,Guo2020,Wu2019}, by jointly optimizing the active and passive beamforming. RIS can also be used to improve the secrecy rate even when the channel responses of the legitimate receiver and the eavesdroppers are highly correlated \cite{Chen2019}. Moreover, thanks to its ability to suppress the interference, RIS is also widely exploited in cognitive radio networks \cite{yuan2019intelligent,xu2020resource} and non-orthogonal multiple access \cite{yang2019intelligent}. Besides, by embedding energy-intensive RF circuits and signal processing units in the surface, RIS can work as a transmitter or receiver \cite{Alexandropoulos2020a,Huang2020H}, also refer to as large intelligent surface (LIS). The LIS is attached with some RF chains to realize the hybrid A/D beamforming \cite{Shlezinger2020} as an efficient realization of massive antenna arrays for wireless communications. The LIS proposed in \cite{Hu2018,Dardari2020} uses the whole contiguous surface of electromagnetically active material for transmitting and receiving. The advantage of the proposed LIS is that it uses the surface to form a spatially continuous transceiver aperture which can achieve higher spatial resolution.

Generally, RIS with passive loads (positive resistances) is supposed to be passive or nearly-passive that no power is needed when it manipulates the EM signal impinging upon it with the fixed reflecting coefficients. \correct{}{To cater for variable wireless environment, some necessary control and switch circuits are deployed at RIS, enabling each RE to reconfigure its reflection coefficient accordingly \cite{Renzo2020}. The minimal power to support such circuits can be tens of microwatt per RE \cite{kaina2014shaping}. Nevertheless, such a passive RIS-aided system is specifically constrained by the double-fading attenuation, since the reflected signals have to go through a cascaded channel, composed of the transmitter-RIS and the RIS-receiver sub-channels \cite{He2020Cascaded}. Notice that the path loss of the RIS-aided link is proportional to $d_f^{\eta_f} d_b^{\eta_b} \beta_f\beta_b$, where $d_{f}$, $\eta_{f}$, and $\beta_{f}$ are the distance between the transmitter and the RIS, the pathloss exponent and the pathloss measured at the reference distance for the transmitter-RIS sub-channel, respectively, while $d_{b}$, $\eta_{b}$, and $\beta_{b}$ are the corresponding ones for the RIS-receiver sub-channel \cite{Griffin2009,Oezdogan2019}. In order to enhance the RIS-aided link to a reasonable level, passive RIS requires a large number of REs, which introduces a large surface size. However, the maximum number of REs in a passive RIS is also constrained by the power budget at the RIS, as each RE still needs extra power to support the aforementioned circuit operations \cite{Huang2019}, which hinders the capacity to assist the existing communication.}

In this paper, we propose an active RIS in which each RE is supported by a set of active-load impedances. Compared with its passive counterpart, the active RIS is likely to an active reflector, which directly reflects the incident signal with the power amplification in the EM level, and it still takes the advantage of RIS that no complex and power-hungry RF chain components are needed. In practice, the active-load RE can be realized by exploiting negative resistance components, such as tunnel diodes, to convert DC bias power into RF power, thus amplifying the incident signal without significantly affecting the low power budget requirement \cite{Bousquet20124,Amato2018}. Provided that these active-load devices are able to harvest the energy from the external environment, they have the potential in realizing the energy neutrality \cite{Kimionis2014b}. Such an active-load RE has been recently introduced into backscatter communication (BSC), another kind of reflective radio technologies, to enhance the backscatter link \cite{Amato2018,Khaledian2019,Amato2018a,Varshney2019TunnelScatter,Kimionis2014a}. The active-load RE helps to achieve orders of magnitude improvement in the BSC communication range \cite{Amato2018a,Amato2018,Kimionis2014a,Varshney2019TunnelScatter}.

As RIS shares the same EM scattering principle with BSC, this motivates us to introduce active loads into the RIS design. In this way, an active RIS tackles the double-fading attenuation by not only increasing the number of REs but also amplifying the incident signals at each RE. With the ability to amplify the incident signal, fewer number of REs are required to achieve a target SNR at the receiver. The trick of active RIS is that after the incident signal that goes through the transmitter-RIS sub-channel, it is usually of weak signal power, and thus can be easily amplified with a considerable gain at the cost of lower power consumption. Compared to the gain provided with the increase in the number of RE as the case in the passive RIS, the gain due to the active RIS power amplification is more straightforward and efficient. Thus, the physical size of the RIS can be shrunk, making it suitable to the space-limited scenarios in which the large passive RIS cannot be applied. Furthermore, it offers the flexibility to reconfigure the wireless propagation environment by optimizing the amplitudes of the reflection coefficients instead of just phases, resulting in a more spectrum and energy-efficient communication.

\correct{}{In particular, we apply the proposed active RIS to a single-input multiple-output (SIMO) communication system, where the RIS assists the transmission from a single-antenna transmitter (Tx) to a multiple-antenna receiver (Rx). This scenario usually takes place when the mobile user uploads its information to the base station via a uplink channel.  With the help of active RIS, the power-limited mobile user can use less transmit power to realize the same performance.} The active RIS however brings a special challenge into the joint design of receive beamforming and RIS reflect beamforming (reflection coefficients), as the proposed RIS can also accidentally amplify the noise at the RIS, resulting in correlated noises at the Rx. Conventionally, such correlated noises coming from passive RIS is usually ignored, since the passive REs adopted in most existing RIS setups just introduce a phase shift in the incident signal without amplifying it. Thus, the REs' reflecting coefficients (reflect beamforming) of the active RIS have to be redesigned to balance the two conflicting goals, i.e., the received signal power maximization and the noise power minimization. In addition, the receive beamforming design at the Rx also needs to consider the correlated noise effect when maximizing the received SNR.


In addition, the active RIS enables itself to strengthen the RIS-aided link by either increasing the number of active REs or amplifying the incident signal. With the increase of the number of REs, the hardware power consumption for support these REs' circuits will inevitably increase. As a result, less power will be allocated to the signal amplification, provided that the power budget at the active RIS is constrained. This may conversely degrade the strength of the reflected signal. It is obvious that for a given power budget at the active RIS, there exists a tradeoff between the number of active REs and the amplification power at each element. Hence, it is worth investigating how many active REs are needed to achieve the optimal SNR performance at the Rx.

\correct{}{Although the active RIS proposed in this paper is capable of amplifying the incident signals, it is quite different from the amplify-and-forward (AF) relay when it comes to the practical issues. Our proposed active RIS basically inherits the property and hardware structure of the conventional passive RIS except that the reconfigurable passive load impedances are replaced with the active load impedances. Though the active RIS needs additional power consumption to support its active load impedances, its basic operation mechanism is still the same as the other RIS, which directly reflects the incident signal with the desired adjustment in the EM level. Thus, the active RIS exploits the EM scattering principles to amplify the signals in the air without the reception. But for the AF relay, it generally needs bulkier and power-consuming RF chains to receive the signal first and then transmit it with the amplification, which usually takes place at the baseband level and takes two time slots to complete the amplify-and-forward processing. Even though the AF works in the  full-duplex mode, it will increase hardware complexity to mitigate the self-interference. In addition, the active RIS not only amplifies the incident signals but also reconfigures their phases to make the desired signal added constructively at the receiver, while the AF relay directly amplifies the received signal without the phase modification. Therefore, in the AF relay-aided system, the additional phase adjustment is needed at the receiver. For more discussions on the comparison between the RIS and AF relay can be found in \cite{DiRenzo2020}.}

To the best of our knowledge, this is the first work that introduces the active load into the RIS-aided communications. The main contributions of this paper are summarized as follows:

\begin{itemize}

  \item First, we introduce an active RIS into a SIMO communication system to alleviate the double-fading attenuation. In this model, we compare the active RIS with the conventional passive RIS, by formulating the respective SNR maximization problems under the same power budget at the RIS.

  \item Second, we solve the non-convex SNR maximization problem via an iterative algorithm which alternatively optimizes the receive beamforming at the Rx and the reflecting coefficient matrix at the RIS. In particular, the receive beamforming vector is obtained with a closed-form solution based on linear minimum-mean-square-error (MMSE) detection, while the reflect beamforming vector is obtained via the sequential convex approximation (SCA).

  \item Third, we study how many active REs are needed to optimize the SNR at the Rx for the specific case where the related channels are line-of-sight (LOS). Based on the channel model, we derive the upper bound of the received SNR with respect to the number of REs and find the optimal number for REs. The results can also be extended into a more general channel model, which is verified via numerical results.

  \item Finally, simulation results are provided to demonstrate that by properly designing the number of REs, the proposed active RIS outperforms the conventional passive RIS for a given power budget. Thus, the active RIS can help design a more spectrum- and energy-efficient wireless communication.
\end{itemize}

 The rest of this paper is organized as follows. In Section \ref{sec:SystemModel}, we present the preliminary of active load and introduce the proposed active RIS-aided SIMO system model. In Section \ref{sec:problemformulation}, we formulate the SNR maximization problem for both passive and active RIS-aided SIMO system.  In Section \ref{sec:Solution}, we propose an alternating optimization algorithm to solve the proposed SNR maximization problem for the active RIS design. In Section \ref{sec:M}, we study the impact of the number of REs on the performance of the proposed system. In Section \ref{sec:Simulation}, numerical results are presented for performance evaluations. Finally, the paper is concluded in Section \ref{sec:Conclusion}.

  The major notations in this paper are listed as follows: The lowercase, boldface lowercase, and boldface uppercase letters such as  $x$, $\mathbf{x}$, and $\mathbf{X}$ denote the scalar, vector, and matrix, respectively. $|x|$ denotes the absolute value of $x$, and $|\mathbf{x}|$ denotes the vector element-wise absolute value of $\mathbf{x}$. $\|\mathbf{x}\|$ denotes the norm of vector $\mathbf{x}$. ${\cal{CN}}(\mu, \sigma^2)$ denotes the complex Gaussian distribution with mean $\mu$ and variance $\sigma^2$. $\mathbb{E}[\cdot]$ denotes the statistical expectation. $x^{\ast}$ denotes the conjugate of $x$. $\mathbf{X}^{\mathrm{T}}$ and $\mathbf{X}^{\mathrm{H}}$ denotes the transpose and conjugate transpose of matrix $\mathbf{X}$, respectively.

\section{System Model}\label{sec:SystemModel}
In this paper, we consider an active RIS-aided SIMO system, which consists of a single-antenna transmitter (Tx), an active RIS equipped with $M$ REs, and a receiver (Rx) equipped with $N$ antennas. The active RIS dynamically adjusts the reflecting coefficients at the REs, also referred to as reflect beamforming, to reconfigure the incident signal with the desired phase shift and power amplification. Thus, the Tx can transmit its information via two links, the direct link and the RIS-aided link. In particular, each RE is assumed to be supported by the active load or negative resistance, which enables the REs to shift the phase as well as to amplify the amplitude of the incident signal \cite{Khaledian2019,Kimionis2014a}. However, the signal amplification at the active RIS also inevitably amplifies the received noise at the RIS, leading to the increase of noise power at the Rx. In order to maximize the signal power while suppressing the correlated noise power at the Rx, the system has to optimize its reflecting coefficient matrix at the RIS and the receive beamforming at the Rx. Before we go through the details of our proposed system, we first have a brief investigation on how the active RIS works.

\subsection{Preliminary}
Reflective radio is based on the EM scattering principles, by which each RE leverages the variations of the reflecting coefficients determined by the impedances of the RE antenna and its input load. Specifically, the value of the reflecting coefficient depends on the following equation,
\begin{equation}\label{eq:reflect}
  \Gamma = \frac{Z_{L}-Z_{A}^{*}}{Z_{L}+Z_{A}},
\end{equation}
where $Z_{L}$ and $Z_{A}$ represent the load and the antenna impedances, respectively. In general, to realize ultra-low-cost and ultra-low-power communication, the passive RIS takes the advantages of the passive load, where the reflecting coefficient is limited by $|\Gamma|^2\leq1$. In our proposed active RIS, the RE works as a reflection amplifier, which allows the reflecting coefficient greater than unity, i.e., $|\Gamma|^2\geq1$. In particular, the RE is characterized by an active load (or negative resistance) impedance at the cost of a certain amount of biasing voltage \cite{Amato2018}, and thus its impedance is shown as follows,
\begin{equation}\label{eq:ActiveLoad}
  Z_{L}=-R_{L}+jX_{L},\quad R_{L}>0,
\end{equation}
where the real part $-R_{L}$ and imaginary part $X_L$ of the load impedance represent the resistance and the reactance, respectively. Meanwhile, the antenna impedance is characterized by the regular passive load, shown as $Z_{A}=R_{A}+jX_A,~R_{A}>0$, where $R_{A}$ and $X_A$ of the antenna impedance represent the resistance and the reactance, respectively. Then, the amplitude of the RE can be found to have the following form
\begin{equation}\label{eq:AL_RE}
    |\Gamma|^2 =\frac{\left(R_L+R_A\right)^2+\left(X_L+X_A\right)^2}{\left(R_L-R_A\right)^2+\left(X_L+X_A\right)^2}>1,
\end{equation}
which enables the RE to serves as a low-power reflection amplifier.

\begin{figure}[t]
  \centering\includegraphics[width=.6\columnwidth]{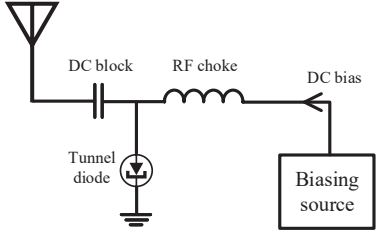} %
  \caption{A simple block diagram of a tunneling reflection amplifier \cite{Amato2018}.}\label{fig:RF}
\end{figure}

{The low-power reflection amplifier can realize the signal amplification with the tunnel diode circuit, which is illustrated in the simple block diagram Fig.~\ref{fig:RF}. The DC-block capacitor here is used to block the low-frequency component from the input RF signal, while the RF choke inductor is used to prevent the high-frequency component from the biasing source. With the suitable DC biasing source to the tunnel diode, it can work at a bias point in the negative resistance region. Then, by matching the input impedance of the tunnel diode to a desired value, the stability in the reflection gain is thus achieved \cite{Khaledian2017}, resulting in the signal amplification in the EM level. Notice that the energy conservation law still holds for the application of active load, as the amplification is supported by a biasing source, which converts the direct current power into the RF power.}

This tunnel diode based low-power reflection amplifier is still in its infancy, and some limitations need to take care of. The lower reflection amplifier is sensitive to the input power of the incident signal, as the low DC bias source cannot support the high amplification gain in the high input power region \cite{Varshney2019TunnelScatter}. In addition, due to the nonlinear property of tunnel diode, there exists nonlinear operation region where the amplification gain is nonlinear but the phase is almost perfectly preserved. Therefore, the reflection amplifier should be properly deployed at the area where the power of the attenuated incident signal is in its linear operation region \cite{Lee2017}. Besides, even when the reflection amplifier operates in the nonlinear region, it can still assist the signal with the phase modulation.

\correct{}{Furthermore, to reconfigure the incident signal with multi-state amplitude and phase variations, one potential solution is to preset a set of tunnel diodes with properly tuned and biased to present the desired reflecting coefficients \cite{Amato2018a}.
Specifically, by rearranging the reflection coefficient $\Gamma$ from~\eqref{eq:reflect}, a set of complex load impedances is yielded by
\begin{equation}\label{eq:Load_impedance}
  Z_{L} = \frac{\Gamma Z_A+Z^{*}_A}{1-\Gamma}.
\end{equation}
The modulating load impedances found by \eqref{eq:Load_impedance} can be obtained by tuning the inductor, the capacitor and the biased tunnel diodes. Then, to realize the amplitudes and phases change at each RE, the efficient controller design of quadrature amplitude modulation (QAM) in BSC \cite{Thomas2012Quadrature} can be exploited. Such a controller design is also common in the passive RIS to implement the discreet phase shift \cite{Wu2020}. For a given combination of the active loads, the maximum amplification gain that a RE can provide depends on the reflecting coefficient with the maximum amplitude created by one of the active loads.}

\correct{}{In practice, several issues need to be considered to implement the active RIS. First, the reflection amplitude is phase-dependent due to the circuit constraint, leading to additional constraint between the reflection amplitude and phase \cite{Abeywickrama2020,jung2019optimality}. This dependency can be attenuated with the improvement over the effective resistance of the load impedance generating circuit \cite{Abeywickrama2020}. Second, the variation of the load impedance is discreet with finite states due to the hardware limitations and the limited control link \cite{jung2019optimality,Wu2020}. It has been shown that with the proposed algorithm in \cite{jung2019optimality,Wu2020}, the RIS with 3 bits-resolution RE achieves almost the same performance with the continuous RE. As we aim to show our main design  methodology for the proposed active RIS and get some insights from its upper bound performance, we assume that the amplitude and the phase of each RE can be tuned independently. Furthermore, the amplitude can be continuously varied with a maximum value constraint and the phase can also be continuously shifted in $\left[0, 2\pi\right)$ in the remaining of this paper. The study of active RIS with these more practical constraints will be left for future work.}





\subsection{Channel Model}
\begin{figure}[t]
  \centering\includegraphics[width=.8\columnwidth]{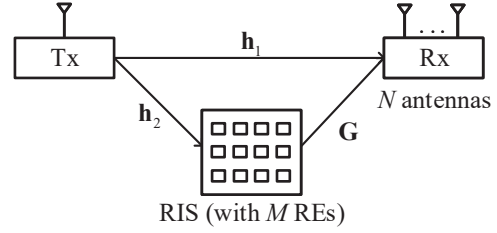} %
  \caption{SIMO system with an active RIS.}\label{fig:systemmodel}
\end{figure}

As shown in Fig. \ref{fig:systemmodel}, the direct-link channel from the Tx to the Rx is denoted by $\mathbf{h}_{1}\in\mathbb{C}^{N \times 1}$. The RIS-aided link is composed of a forward channel from the Tx to the RIS and a backward channel from the RIS to the Rx, denoted by $\mathbf{h}_{2}\in\mathbb{C}^{M \times 1}$ and $\mathbf{G}\in\mathbb{C}^{N \times M}$, respectively. We assume that all the channels in the system undergo independent block fading, and the instantaneous channel state information (CSI) for all links is assumed to be available to the Rx and the RIS at each fading block. In practice, the RIS-aided CSI can be estimated with the compressive sensing technique when the RIS-aided cascaded channels are typically sparse \cite{chen2019channel}. Also, some active sensors can be deployed at the RIS to facilitate the channel estimation \cite{Alexandropoulos2020,Taha2019}.

The RIS is expected to intelligently reconfigure the wireless channel based on the existing CSI by varying its reflecting coefficient matrix (i.e., reflect beamforming), $\mathbf{\Phi}=\mathrm{diag}\left(\phi_1,...,\phi_M\right)\in\mathbb{C}^{M\times M}$, which is diagonal.  Different from the most existing RIS designs, we herein consider that the RIS can be supported by an external power source, and thus the REs in the active RIS can exploit the active loads to amplify the incident signal to alleviate the double-fading attenuation.  In particular, the reflecting coefficient of the $m$-th RE at the RIS is denoted by $\phi_m=a_m e^{j\theta_m}, m =1,...,M$, where $a_m$ and $\theta_m$ represent the amplitude and the phase, and $a_m$ can be greater than $1$ with active load. {As aforementioned, we assume that the amplitude and the phase can be continuously varies within the intervals $a_m\in[0, a_{m,\max}], a_{m,\max}\geq 1$ and $\theta_m \in [0, 2\pi)$, respectively.}


\subsection{Signal Model}
 Let us denote the transmit signal at the Tx by $s(n)$. We assume $s(n)$ is an independent complex Gaussian random variable with zero mean and unit variance. Denote the transmit power as $p_\mathrm{t}$. The transmit signal reaches the Rx via both the direct link and the RIS-aided link. Thus, the received signal at the Rx is written as
\begin{align}
\mathbf{y}(n) &= \underbrace{\sqrt{p_\mathrm{t}}\mathbf{h}_{1}s(n)}_{\text{direct link}}+\underbrace{\mathbf{G}\mathbf{\Phi}\left(\sqrt{p_\mathrm{t}}\mathbf{h}_2  s(n)+\mathbf{z}_2(n) \right)}_{\text{RIS-aided link}}+\mathbf{z}_1(n) \nonumber \\
&= \sqrt{p_\mathrm{t}}\left(\mathbf{h}_{1}+ \mathbf{G}\mathbf{\Phi}\mathbf{h}_2\right)s(n)+ \mathbf{G}\mathbf{\Phi}\mathbf{z}_2(n) +\mathbf{z}_1(n),\label{eq:ReceivedSingal}
\end{align}
where $\mathbf{z}_1(n)\in\mathbb{C}^{N\times 1}$ and $\mathbf{z}_2(n)\in\mathbb{C}^{M\times 1}$ represent the thermal noise at the Rx and the RIS, which are distributed as $\mathcal{CN}\left(\mathbf{0},\sigma^2_1\mathbf{I}_{N}\right)$ and $\mathcal{CN}\left(\mathbf{0},\sigma^2_2\mathbf{I}_{M}\right)$, respectively. Notice that the second term in~\eqref{eq:ReceivedSingal} is usually ignored in most literature, since it is considered as a relatively small interference compared to the noise power at the Rx after the reflection operations and going through the backward channel. However, as the active RIS has the ability to amplify the incident signal, the noise introduced at the RIS can be no longer ignored. In addition, it allows us to quantify the SNR loss in the RIS-aided system more accurately by considering the correlated noise effects introduced at the RIS.



After receiving the signal, the multiple-antenna Rx exploits the linear receive beamforming to extract the useful signal for decoding $s(n)$. Thus, the received signal vector $\mathbf{y}(n)$ is transformed to a scalar through receive beamforming as follows
\begin{align}\label{eq:ReceiveBeamforming}
r(n) &= \mathbf{w}^\mathrm{H} \mathbf{y}(n) \nonumber \\
&=\mathbf{w}^\mathrm{H}\!\!\sqrt{p_\mathrm{t}}\left(\mathbf{h}_{1}\!\!+\! \mathbf{G}\mathbf{\Phi}\mathbf{h}_2\right)s(n)\!\!+\! \mathbf{w}^\mathrm{H} \mathbf{G}\mathbf{\Phi}\mathbf{z}_2(n)\! \!+\!\mathbf{w}^\mathrm{H}\mathbf{z}_1(n),
\end{align}
where $\mathbf{w}\in\mathbb{C}^{N\times 1}$ is the receive beamforming vector. The SNR for decoding $s(n)$ is thus written as,
\begin{align}\label{eq:SINR}
  \gamma_s = \frac{p_\mathrm{t}\left|\mathbf{w}^\mathrm{H}\left(\mathbf{h}_1+\mathbf{G}\mathbf{\Phi}\mathbf{h}_2\right)\right|^2}{\sigma_2^2\left\|\mathbf{w}^\mathrm{H}\mathbf{G\Phi}\right\|^2+\sigma_1^2\left\|\mathbf{w}\right\|^2}.
\end{align}
Notice that from~\eqref{eq:SINR}, although the active RIS can improve the received signal power by reconfiguring its reflecting coefficient matrix $\mathbf{\Phi}$, the noise power is also increased together with $\mathbf{\Phi}$, which implies that the active RIS cannot obtain the optimal performance for the received SNR by simply increasing the amplification power. With the SNR expression, the achievable rate for the active RIS-aided SIMO system is written as
\begin{equation}
  R_s = \log_2\left(1+\gamma_s\right).
\end{equation}

\subsection{Power Consumption at the RIS}
In this subsection, we propose two general models which describe the power consumption at the passive and the active RIS, respectively. {The passive RIS power consumption is mainly due to the switch and control circuit at the RE \cite{Huang2019,Renzo2020}. Assuming the passive RIS is equipped with $M$ identical passive RE, the power dissipated at the passive RIS is written as
\begin{equation}\label{eq:Power_passiveRIS}
P_\mathrm{RIS,p} = M P_{\mathrm{c}},
\end{equation}
where $P_\mathrm{c}$ represents the power consumption of the switch and control circuit at each RE. }

{To model the power consumption of the active RIS, we first investigate the power dissipated at each RE with active loads. In the active RIS, the RE forces the tunnel diode (or other electronics with negative resistance property) to essentially work as a reflection amplifier with the suitable DC biasing source. In practice, the reflection amplifier power consumption $P_\mathrm{RE}$ depends on many factors including the specific hardware implementation, DC power, load characteristics, and output power, which is difficult to precisely characterize. To obtain a sufficiently accurate-and-tractable, but practical, power consumption model, we characterize the reflection amplifier power consumption with the classical power amplifier model in \cite{Joung2015}, shown as follows:
\begin{equation}\label{eq:REActiveRIS}
P_{\mathrm{RE},m} = \mathsf{OPI} + \mathsf{OPD}(p_{\mathrm{out},m}),
\end{equation}
where $p_{\mathrm{out},m}$ denotes the output signal power at the $m$-th RE, the first component $\mathsf{OPI}$ is the output power independent term (OPI), and the second component $\mathsf{OPD}(p_{\mathrm{out},m})$ is the output power dependent term (OPD). For the proposed RE, the OPI term includes the switch and control circuit power consumption and the DC biasing power consumption at each RE, which is written as
\begin{equation}\label{eq:REOPI}
\mathsf{OPI} = P_\mathrm{c}+P_\mathrm{DC}.
\end{equation}
While the OPD term can be empirically expressed with a linear model as
\begin{equation}\label{eq:REOPD}
\mathsf{OPD}(p_{\mathrm{out},m}) = \xi p_{\mathrm{out},m},
\end{equation}
where $\xi\triangleq \upsilon^{-1}$ with  $\upsilon$ being the amplifier efficiency, and the output power is related to the incident signal power $p_{\mathrm{in}, m}$ in the form of
\begin{equation}\label{eq:REoutput}
  p_{\mathrm{out}, m} = a_m^2p_{\mathrm{in}, m},
\end{equation}
where we have $p_{\mathrm{in}, m} = |h_{2,m}|^2 p_t  + \sigma_2^2$. It should be noted that we herein assume the RE reflection amplifier operates in its linear region where the output power is linearly increasing with the input power. This assumption is valid for the tunnel diode-based reflection amplifier when the input power is in its operation range \cite{Lee2017}. In fact, as the transmit signal from the Tx goes through the Tx-RIS channel attenuation before it reaches each RE, the incident signal power is so weak that the reflection amplifier can obtain a high amplification gain with the low output power. The output amplification power typically takes up a small portion of the power consumption compared to the DC power consumption \cite{Amato2018}. For this reason, most existing literatures estimate the total power consumption of this reflection amplifier with the DC power consumption. However, in our work, we do not ignore the OPD term to model the power consumption more accurately.}

{Considering the active RIS is equipped with $M$ identical active REs, the total power consumption is thus given by
\begin{equation}\label{eq:ActiveRISPower}
   P_\mathrm{RIS,a} \triangleq M (P_\mathrm{c} + P_\mathrm{DC})+ \xi p_\mathrm{out},
\end{equation}
where $p_\mathrm{out}$ is the output power of the active RIS, and is given by $p_\mathrm{out}=p_\mathrm{t}\left\|\mathbf{\Phi}\mathbf{h}_2 \right\|^2+\sigma_2^2\left\|\mathbf{\Phi}\mathbf{I}_{M}\right\|^2$. Although such an active RIS needs additional power consumption to support the active load, the active RIS enables each RE to individually modulate the incident signal in the EM-level as the case of passive RIS, but with amplifying it. Thus, the RE in the active RIS realizes the signal amplification in a lightweight manner without the presence of the complex and power-hungry RF chain components, which reduces plenty of power consumption. It has also been shown that the power consumption of reflection amplifiers has been decreasing to the microwatt level during the recent years, and tunnel diode-based reflection amplifiers can be a natural step towards less power-hungry electronics \cite{Amato2018}. With the active RIS, the system can benefit from  not only the constructive signal combination at the Rx but also the power amplification.}

In particular, for a given power budget $P_\mathrm{RIS,a}$, the active RIS is allowed to allocate the remaining power to amplify the incident signal with active loads, after supplying the hardware power consumption for $M$ REs. With the active load, the amplitude interval $a_m\in\left[0,1\right]$ can be extended. For a predetermined maximum amplitude $a_{m,\max}$ that the active load can provide, each active-load RE has the following amplitude constraint
\begin{equation}\label{eq:maxBounded}
  a_m \leq a_{m,\max},\quad \forall m.
\end{equation}

Moreover, the amplification power is limited due to the overall power budget at the RIS. When designing the RIS reflecting coefficient matrix $\mathbf{\Phi}$, the amplification power constraint is considered as follows,
\begin{equation}\label{eq:RIS_power_constraint}
  p_\mathrm{t}\left\|\mathbf{\Phi}\mathbf{h}_2 \right\|^2+\sigma_2^2\left\|\mathbf{\Phi}\mathbf{I}_{M}\right\|^2 \leq \upsilon \left(P_\mathrm{RIS,a} - M \left(P_\mathrm{c}+P_\mathrm{DC}\right)\right).
\end{equation}
Therefore, for an active RIS with $M$ REs, the amplification power budget is defined by
\begin{equation}\label{eq:M_PA}
  P_\mathrm{out} \triangleq \upsilon \left(P_\mathrm{RIS,a} - M \left(P_\mathrm{c}+P_\mathrm{DC}\right)\right).
\end{equation}
Notice that this constraint is introduced only in the active RIS-aided communication system, and it is not suitable for the passive RIS, as the amplitude of the passive RE admits $a_m \in \left[0,1\right]$ which does not have the ability to amplify the incident signal. Such passive RIS design problems have been widely discussed in \cite{Huang2019,Wu2019,Chen2019}.

\section{Problem Formulation}\label{sec:problemformulation}
\subsection{Passive RIS for SIMO Systems}
Before investigating the performance of the active RIS-aided system, we first study the classic passive RIS design problem in such an SIMO system for comparison. As aforementioned, the conventional RIS is equipped with $M$ passive REs, each of which is supported by controllable positive resistances, i.e., passive loads, and consumes $P_\mathrm{c}$ circuit power for control operation. Note that the passive RE herein is considered to only adjust the phase of the incident signal without power amplification, and thus the noise introduced at the RIS in \eqref{eq:ReceivedSingal} is neglected as compared to the noise power at the Rx. Therefore, the received SNR at the Rx is written as
\begin{equation}\label{eq:SINR_passive}
  \gamma_s = \frac{p_\mathrm{t}\left|\mathbf{w}^\mathrm{H}\left(\mathbf{h}_1+\mathbf{G}\mathbf{\Phi}\mathbf{h}_2\right)\right|^2}{\sigma^2_1}.
\end{equation}
For a given $\mathbf{\Phi}$, it is well-known that the optimal receive beamforming design scheme to maximize the SNR is maximal-ratio-combining (MRC), of which the receive beamforming is shown as
\begin{equation}\label{eq:RxBeamforming_passive}
  \mathbf{w} = \frac{\mathbf{h}_1+\mathbf{G}\mathbf{\Phi}\mathbf{h}_2}{\left\|\mathbf{h}_1+\mathbf{G}\mathbf{\Phi}\mathbf{h}_2\right\|}.
\end{equation}
With the MRC beamforming design, the SNR maximization problem is equivalent to the following channel power gain maximization problem:
\begin{subequations}
  \begin{align}
  \max_{\mathbf{\Phi}}\quad &\left\|\mathbf{h}_1+\mathbf{G}\mathbf{\Phi}\mathbf{h}_2\right\|^2 \\
  \mathrm{s.t.} \quad&  a_{m} = 1, \forall m \label{eq:unit1}\\
  & 0\leq \theta_m<2 \pi, \forall m. \label{eq:unit2}
\end{align}
\end{subequations}
{The above problem has been well investigated in \cite{Wu2019} and efficiently solved by alternating optimization technique.} In addition, one of the important results for the passive RIS-aided system is that the received SNR increases with $M$. Therefore, for a limited RIS power budget $P_\mathrm{RIS}$, the optimal SNR is obtained when the number of passive REs is maximized with $M^{*} = \lfloor P_\mathrm{RIS}/P_\mathrm{c} \rfloor$.

\subsection{Active RIS for SIMO Systems}
Different from the passive RIS design, the REs in the active RIS will introduce the amplitude constraint~\eqref{eq:maxBounded} and the amplification power constraint~\eqref{eq:RIS_power_constraint} instead of the amplitude constraint shown in~\eqref{eq:unit1} and~\eqref{eq:unit2}. In particular, we aim to optimize the reflecting coefficient matrix at the active RIS and the receive beamforming at the Rx to maximize the received SNR, subject to the RIS power budget constraint. The problem is formulated as
\begin{subequations}
\begin{align}
 \textbf{P1}:\quad \max_{\mathbf{\Phi},\mathbf{w}} & \quad \frac{p_\mathrm{t}\left|\mathbf{w}^\mathrm{H}\left(\mathbf{h}_1+\mathbf{G}\mathbf{\Phi}\mathbf{h}_2\right)\right|^2}{\sigma_2^2\left\|\mathbf{w}^\mathrm{H}\mathbf{G\Phi}\right\|^2+\sigma_1^2\left\|\mathbf{w}\right\|^2} \label{eq:Obj_1}\\
 \mathrm{s.t.}  &\quad p_\mathrm{t}\left\|\mathbf{\Phi}\mathbf{h}_2 \right\|^2+\sigma_2^2\left\|\mathbf{\Phi}\mathbf{I}_{M}\right\|^2 \leq P_\mathrm{out}, \label{eq:P1C1} \\
 &\quad a_m\leq a_{m,\max},~\forall m, \label{eq:P1C3} \\
 & \quad\|\mathbf{w}\|^2 = 1, \label{eq:P1C2}
\end{align}
\end{subequations}
where \eqref{eq:P1C2} represents the normalization constraint for the receive beamforming. It is worth mentioning that the problem is formulated under the assumption that the number of REs $M$ is fixed. In fact, $M$ is also an important optimization variable, as the RIS-aided signal can be strengthened with the increase in both $M$ and $P_\mathrm{out}$. In general, more REs can provide better performance in SNR as presented in \cite{Liang2019}. However, according to~\eqref{eq:M_PA}, with more REs, less power is left to amplify the incident signal, which may conversely decrease the SNR performance. Obviously, there exists a novel tradeoff between the number of REs and the power amplification in the active RIS design, which motivates us to discuss the optimization on $M$ in the later section.

For any given reflecting coefficient matrix $\mathbf{\Phi}$, it is well-known that the linear minimum-mean-square-error (MMSE) detector is the optimal receive beamforming to problem (P1) as it can cope with the increased interference caused by the active RIS noise amplification. The MMSE-based receive beamforming is written as
\begin{equation}\label{eq:RxBeamforming_active}
  \mathbf{w}^{*} = \frac{\left(\mathbf{h}\mathbf{h}^\mathrm{H}+\frac{\sigma_2^2}{p_\mathrm{t}}\mathbf{G}\mathbf{\Phi}\mathbf{\Phi}^\mathrm{H}\mathbf{G}^\mathrm{H}+\frac{\sigma_1^2}{p_\mathrm{t}}\mathbf{I}_{N}\right)^{-1}\mathbf{h}}{\left\|\left(\mathbf{h}\mathbf{h}^\mathrm{H}+\frac{\sigma_2^2}{p_\mathrm{t}}\mathbf{G}\mathbf{\Phi}\mathbf{\Phi}^\mathrm{H}\mathbf{G}^\mathrm{H}+\frac{\sigma_1^2}{p_\mathrm{t}}\mathbf{I}_{N}\right)^{-1}\mathbf{h}\right\|},
\end{equation}
where we have the equivalent channel $\mathbf{h}=\mathbf{h}_1+\mathbf{G}\mathbf{\Phi}\mathbf{h}_2 \in \mathbb{C}^{N\times1}$. With $\mathbf{w}^{*}$, problem (P1) is recast into the following problem
\begin{subequations}
  \begin{align}
   \max_{\mathbf{\Phi}} & \quad \mathbf{h}^\mathrm{H}\left(\frac{\sigma_2^2}{p_\mathrm{t}}\mathbf{G}\mathbf{\Phi}\mathbf{\Phi}^\mathrm{H}\mathbf{G}^\mathrm{H}+\frac{\sigma_1^2}{p_\mathrm{t}}\mathbf{I}_{N}\right)^{-1}\mathbf{h} \label{eq:Obj_12}\\
   \mathrm{s.t.}  &\quad p_\mathrm{t}\left\|\mathbf{\Phi}\mathbf{h}_2 \right\|^2+\sigma_2^2\left\|\mathbf{\Phi}\mathbf{I}_{M}\right\|^2 \leq P_\mathrm{out}. \label{eq:P1C12} \\
   &\quad a_m\leq a_{m,\max},~\forall m, \label{eq:P1C13}
  \end{align}
\end{subequations}
Although these constraints are convex, it is challenging to solve the above problem due to the nonconvex objective~\eqref{eq:Obj_12} where both the equivalent channel and the effective noise, shown in the numerator and denominator, respectively, depend on $\mathbf{\Phi}$.


\section{Alternating Optimization Solution}\label{sec:Solution}
To solve the original problem (P1), we propose an alternating optimization for the RIS-aided system, by which the above problem is solved in two stages iteratively. In the first stage, the reflecting coefficient matrix $\mathbf{\Phi}$ at the RIS is optimized for a given receive beamforming $\mathbf{w}$. In the second stage, the receive beamforming $\mathbf{w}$ at the Rx is optimized for a given reflecting coefficient matrix $\mathbf{\Phi}$. By iteratively updating the optimization results in the two stages, an efficient solution to the original problem can be obtained. Specifically, (P1) is now decoupled into the following two subproblems (P1-A) and (P1-B):
\begin{align}
& \textbf{P1-A}: \max_{\mathbf{\Phi}}~~ \gamma_{s}(\mathbf{\Phi},\mathbf{w}) \nonumber\\
& \qquad\quad\mathrm{s.t.}~~~ p_\mathrm{t}\left\|\mathbf{\Phi}\mathbf{h}_2 \right\|^2+\sigma_2^2\left\|\mathbf{\Phi}\mathbf{I}_M\right\|^2 \leq P_\mathrm{out},\nonumber \\
&\qquad~~~~\qquad a_m\leq a_{m,\max},~\forall m, \nonumber \\
& \textbf{P1-B}: \max_{\mathbf{w}}~~ \gamma_{s}(\mathbf{\Phi},\mathbf{w})\quad \mathrm{s.t.}~~ \|\mathbf{w}\|^2=1. \nonumber
\end{align}

As aforementioned, for any given reflecting coefficient matrix $\mathbf{\Phi}$, the linear MMSE receive beamforming is optimal. Thus, the optimal solution to problem (P1-B) is obtained with \eqref{eq:RxBeamforming_active}.

In the following, we focus on the solution to optimization problem (P1-A). For a given receive beamforming $\mathbf{w}$, the Rx can be equivalently regarded as a single-antenna Rx. Consequently, the direct-link channel is SISO and can be represented by a scalar $h_1=\mathbf{w}^\mathrm{H}\mathbf{h}_1 \in \mathbb{C}^{1\times 1}$, while the backward channel is MISO and can be represented by a vector $\mathbf{g} =\mathbf{w}^\mathrm{H}\mathbf{G} \in\mathbb{C}^{1\times M}$.
Furthermore, let $\mathbf{a}=\left[\phi_1,...,\phi_M\right]^\mathrm{T}$, $\mathbf{b}=\mathrm{diag}\left(\mathbf{h}_2^{\mathrm{H}}\right)\mathbf{g}^{\mathrm{H}}$, $\mathbf{Q} = \mathrm{diag}\left([|g_1|^2,|g_2|^2,...,|g_M|^2]\right)$, and  $\mathbf{F}=p_\mathrm{t}\mathrm{diag}([|h_{21}|^2,\dots,|h_{2M}|^2])+\sigma_2^2\mathbf{I}_M$. Problem (P1-A) can be recast into the following problem with respect to the reflect beamforming vector $\mathbf{a}$,
\begin{subequations}
  \begin{align}
\max_{\mathbf{a}}&  \qquad \frac{p_\mathrm{t}\left| \mathbf{b}^\mathrm{H}\mathbf{a}+h_1 \right|^2}{{{\mathbf{a}^\mathrm{H}\mathbf{Qa}}\sigma _2^2 +  \sigma _1^2}}\label{eq:P1-A-OBJ1}\\
\mathrm{s.t.}&\qquad \mathbf{a}^\mathrm{H}\mathbf{F}\mathbf{a}\leq P_\mathrm{out}. \label{eq:P1-A-C1} \\
&\qquad a_m\leq a_{m,\max},~\forall m \label{eq:P1-A-C2}
\end{align}
\end{subequations}
The above reflect beamforming problem cannot be solved with the conventional RIS optimization algorithm, as the active RIS introduces additional noise term at the denominator of the objective function, resulting in a quadratic fractional programming problem. However, the classical Dinkelbach method cannot be directly applied to solve this fractional programming problem, since both the numerator and the denominator of the objective function are convex. To further proceed this problem, we first investigate the unconstrained problem of (P1-A) to find the phase design strategy. The optimization problem is now simplified as follows,
\begin{align}\label{eq:Obj_RefB_var}
 \max_{\mathbf{a}} \qquad \frac{p_\mathrm{t}\left| \mathbf{b}^\mathrm{H}\mathbf{a}+h_1 \right|^2}{{{\mathbf{a}^\mathrm{H}\mathbf{Qa}} \sigma _2^2 +  \sigma _1^2}},
\end{align}

\proposition\label{prop:ReflectiveBeamforming} Without any constraint, the optimal reflecting coefficients for each RE in the active RIS is
\begin{align}\label{eq:Opt_Ref}
  \phi_m^{\star} =\frac{\sigma _1^2 |h_{2m}|}{\sigma _2^2|h_1||g_m|} e^{j\left(\mathrm{arg}(h_1)-\mathrm{arg}(h_{2m})-\mathrm{arg}(g_{m})\right)}, \forall m,
\end{align}
where $\mathrm{arg}(\cdot)$ represents the argument of a complex number, and the maximum SNR $\gamma_s^{\star}=\frac{{p_\mathrm{t}{{\left| {{h_1}} \right|}^2}}}{{\sigma _1^2}} + \frac{{p_\mathrm{t}{{\left\| {{{\bf{h}}_2}} \right\|}^2}}}{{\sigma _2^2}}$.
\begin{IEEEproof}
This result can be obtained by setting the first derivative of the objective \eqref{eq:Obj_RefB_var} with regard to $\mathbf{a}$ to 0.
\end{IEEEproof}
From~\eqref{eq:Opt_Ref}, two important insights are presented: First, each RE needs to adjust the phase of the incident signal into the phase of $h_1$ to make the reflected signals combined constructively at the Rx; Second, the amplitude of each RE is designed to mitigate the noise power introduced by the active RIS by dividing the backward channel gain $|g_m|$ and to enhance the received SNR by multiplying the forward channel gain $|h_{2m}|$. In addition, we have an important remark with respect to $\gamma_s$

\remark Due to the correlated noise effect, $\gamma_s$ is no longer an increasing function of the amplification power for the active  system. If the RIS is not deployed, the received SNR will be $\gamma_s=\frac{{p_\mathrm{t}{{\left| {{h_1}} \right|}^2}}}{{\sigma _1^2}}$. If the amplitude of the RE $a_m\rightarrow \infty, \forall m$, the received SNR will be $\gamma_s = \frac{{p_\mathrm{t}{{\left\| {{{\bf{h}}_2}} \right\|}^2}}}{{\sigma _2^2}}$.

Based on the result \eqref{eq:Opt_Ref} from solving the above unconstrained problem, we obtain the optimal phase design for the reflect beamforming without any constraint as follows,
\begin{equation}\label{eq:Opt_phase}
  \theta_m = \mathrm{arg}(h_1)-\mathrm{arg}(h_{2m})-\mathrm{arg}(g_{m}),~\forall m.
\end{equation}

For the original problem, the optimal phase design still holds, due to the fact that the noise power in the denominator of \eqref{eq:P1-A-OBJ1} and the amplification power in \eqref{eq:P1-A-C1} and \eqref{eq:P1-A-C2} are independent of the phase value of each RE, while the phase design in \eqref{eq:Opt_phase} is optimal to maximize the objective \eqref{eq:P1-A-OBJ1}. Therefore, we just need to optimize the amplitude of each RE in the following. This helps to reduce the computational complexity of the proposed algorithm, as it only deals with the real-number variables. After applying the phase design \eqref{eq:Opt_phase}, problem (P1-A) can be rewritten as
\begin{subequations}
  \begin{align}
  \max_{ \bar{\mathbf{a}}} & \qquad \frac{p_\mathrm{t}\left|\bar{\mathbf{b}}^\mathrm{T}\bar{\mathbf{a}}+|h_1|\right|^2}{\bar{\mathbf{a}}^\mathrm{T}{\mathbf{Q}}\bar{\mathbf{a}}\sigma_2^2+\sigma_1^2}\label{eq:P1ASObj}\\
  \mathrm{s.t.} & \qquad \bar{\mathbf{a}}^\mathrm{T}{\mathbf{F}}\bar{\mathbf{a}} \leq P_\mathrm{out},\label{eq:P1ASC1}\\
  & \qquad a_m\leq a_{m,\max},~\forall m,\label{eq:P1ASC2}
\end{align}
\end{subequations}
where $ \bar{\mathbf{a}}=  |\mathbf{a}|$ and $ \bar{\mathbf{b}}= |\mathbf{b}|$.  To solve this problem, we first equivalently converts the problem into a more trackable form by introducing several auxiliary variables. Then, we exploit the sequential convex approximation (SCA) method which iteratively approximates these nonconvex functions with the linear functions, and thus we can obtain the suboptimal reflect beamforming design by efficiently solving a series of convex optimization problems.


The difficulty in solving the maximization problem mainly lies in the objective~\eqref{eq:P1ASObj}. To tackle it, we first rewrite the problem into the equivalent form
\begin{subequations}
\begin{align}
\max_{\tau,\kappa,\bar{\mathbf{a}}} & \qquad \tau \\
\mathrm{s.t.} &\qquad \sqrt{p_\mathrm{t}}\left(\bar{\mathbf{b}}^\mathrm{T}\bar{\mathbf{a}}+|h_1|\right) \geq \sqrt{\tau \kappa} \label{eq:P1ASSC1}\\
  & \qquad \bar{\mathbf{a}}^\mathrm{T}{\mathbf{Q}}\bar{\mathbf{a}}\sigma_2^2+\sigma_1^2 \leq \kappa \label{eq:P1ASSC2}\\
  & \qquad \eqref{eq:P1ASC1} ~\textrm{and}~\eqref{eq:P1ASC2}, \nonumber
\end{align}
\end{subequations}
where the newly introduced auxiliary variables $\tau$ and $\kappa$ represent the received SNR and noise interference power received at the Rx, respectively. Notice that if the equalities $\eqref{eq:P1ASSC1}$ and $\eqref{eq:P1ASSC2}$ hold, the reformulated problem is equivalent to the original problem (P1-A). Notice that~\eqref{eq:P1ASSC2} is a convex quadratic constraint. However, the constraint~\eqref{eq:P1ASSC1} is nonconvex, due to the form which is the square root of the multiplication of two optimization variables.

In the following, we propose a SCA method, which approximates the square root by a convex upper-bound in each iteration. For convenience, we define $\tau_{(t)}$ and $\kappa_{(t)} $ as the iterative optimization variables after the $t$-th step iteration. Then, we exploit the first-order Taylor polynomial of $\sqrt{\tau\kappa}$ around the point $(\tau_{(t)},\kappa_{(t)})$ to approximate the original constraint, and thus we have a convex upper-bound
\begin{align}
\sqrt{\tau\kappa} &\leq \mathcal{G}_{(t)}(\tau,\kappa) \nonumber\\
& \triangleq \sqrt{\tau_{(t)}\kappa_{(t)}}+\frac{1}{2}\left(\frac{\kappa_{(t)}}{\tau_{(t)}}\right)^{1/2}\left(\tau-\tau_{(t)}\right) \nonumber \\
&\quad +\frac{1}{2}\left(\frac{\tau_{(t)}}{\kappa_{(t)}}\right)^{1/2}\left(\kappa-\kappa_{(t)}\right). \label{eq:SCA_linear}
\end{align}
Thus, based on these transformations, the suboptimal solution to the original problem (P1-A) can be obtained with these parameters $(\tau_{(t)},\kappa_{(t)})$ which are iteratively updated by solving the following convex optimization problem
\begin{subequations}\label{eq:P3-SCA}
  \begin{align}
    \textbf{P1-A-SCA:}~~~~ \max_{\tau,\kappa,\bar{\mathbf{a}}} &\quad \tau  \nonumber\\
    \mathrm{s.t.} &\quad  \sqrt{p_\mathrm{t}}\left(\bar{\mathbf{b}}^\mathrm{T}\bar{\mathbf{a}}+|h_1|\right) \geq \mathcal{G}_{(t)}(\tau,\kappa) \\
    &\quad \eqref{eq:P1ASC1},~\eqref{eq:P1ASC2},~ \mathrm{and}~\eqref{eq:P1ASSC2}. \nonumber
  \end{align}
\end{subequations}
Specifically, the initial point $(\tau_{(0)},\kappa_{(0)})$ can be obtained by solving a simple feasible problem subject to the constraints \eqref{eq:P1ASC1} and \eqref{eq:P1ASC2}. Denote the solution to the feasible problem by $\bar{\mathbf{a}}_{(0)}$, and we have the initial $\tau_{(0)} = \frac{p_\mathrm{t}\left( \bar{\mathbf{b}}^\mathrm{T}\bar{\mathbf{a}}_{(0)}+|h_1| \right)^2}{{\bar{\mathbf{a}}_{(0)}^\mathrm{T}\mathbf{Q\bar{a}}_{(0)}}  \sigma _2^2 + \sigma _1^2}$ and $\kappa_{(0)}={\bar{\mathbf{a}}_{(0)}^\mathrm{T}\mathbf{Q\bar{a}}_{(0)}}  \sigma _2^2 + \sigma _1^2$. The details of this algorithm are summarized in Algorithm 1.

\begin{algorithm}[t!]
  \caption{ The proposed SCA-based solution to (P3)} \label{Algorithm_SCA}
  {
  \begin{algorithmic}[1]
  \STATE Initialization: the initial point $(\tau_{(0)},\kappa_{(0)})$. \\
  \REPEAT
  \STATE Update the parametric variables $(\tau_{(t)},\kappa_{(t)})$ by solving the problem (P1-A-SCA) with $(\tau_{(t-1)},\kappa_{(t-1)})$. \\
  \UNTIL{The value of $\tau$ converges.} \\
  \end{algorithmic}
  }
\end{algorithm}

\emph{Converge Analysis of the proposed SCA algorithm:}
In the following, we will analyze the convergence of the proposed SCA-based iterative algorithm. first, we denote $(\tau_{(t)},\kappa_{(t)},\bar{\mathbf{a}}_{(t)})$ as the optimal variables in the $t-1$-th iteration. Due to the linear approximation in \eqref{eq:SCA_linear}, the variable pair $(\tau_{(t)},\kappa_{(t)},\bar{\mathbf{a}}_{(t)})$ is also feasible to the optimization problem at the $t$-th iteration. As $\tau_{(t+1)}$ is the optimal solution to the $t$-th iteration, we thus immediately have $\tau_{(t+1)}\geq \tau_{(t)}$, which means that the proposed SCA algorithm generates a nondecreasing sequence. Furthermore, because of the power constraint, $\tau$ is bounded. Therefore, the convergence of the proposed algorithm is guaranteed. 

{\emph{Converge Analysis of the alternating optimization algorithm:} Overall, for each iteration, the alternating optimization algorithm obtains the optimal solution, which ensures that this algorithm generates a non-decreasing objective function. In addition, due to the power budget constraint \eqref{eq:P1C1}, $\gamma_s$ is bounded above. Therefore, the convergence of the overall algorithm is guaranteed.}

\correct{}{\emph{Complexity Analysis:} Next, we investigate the computational complexity of the proposed alternating optimization algorithm, which consists of the SCA algorithm and MMSE receive beamforming calculation in each iteration. In the SCA algorithm, for each inner iteration, the convex problem (P1-A-SCA) can be equivalently transformed into a convex quadratically constrained quadratic program (QCQP) problem with $\mathsf{m}=M+2$ variables and $\mathsf{n}=M+3$ quadratic constraints. This convex QCQP problem can be solved by the interior-point algorithm with the total arithmetic cost less than $\mathcal{O}\left(\mathsf{m}^{0.5}(\mathsf{m}\mathsf{n}^2+\mathsf{n}^3)\ln\frac{2\mathsf{m}V}{\varepsilon_{q}}\right)$, where $\varepsilon_{q}$ is the prescribed accuracy and $V$ is a constant defined in \cite{nesterov1994interior}. The computational complexity of the MMSE receive beamforming calculation mainly depends on the matrix inversion whose arithmetic cost is $\mathcal{O}(M^3)$. Denote the iteration number for the SCA algorithm and the alternating optimization algorithm by $\mathcal{I}_s$ and $\mathcal{I}_{ao}$, respectively, and the overall algorithm complexity is thus denoted by $\mathcal{O} \left(\mathcal{I}_{ao}\left(\mathcal{I}_s \mathsf{m}^{0.5}(\mathsf{m}\mathsf{n}^2+\mathsf{n}^3)\ln\frac{2\mathsf{m}V}{\varepsilon_{q}}+M^3\right)\right)$, which can be solved in polynomial time. }

\section{How many REs does an active RIS need?}\label{sec:M}
As aforementioned, there exists a tradeoff between the number of REs and the power allocated for signal amplification. In general, more REs provide better performance in SNR as presented in \cite{Liang2019}. However, according to~\eqref{eq:P1C1}, with the increase of the number of REs, less power is allocated to the signal amplification, which may decrease the SNR performance on the contrary.

We investigate the received SNR at the Rx for the proposed active RIS with respect to the number of REs. For simplicity, we assume that the Rx is equipped with a single antenna, i.e., $\mathbf{G}\triangleq\mathbf{g}$. Moreover, we herein ignore the direct link due to the reason that the direct link is independent of the number of REs, and thus the optimal $M$ obtained by maximizing the SNR without the direct link will also be optimal to the case with the direct link. In particular, the line-of-sight (LOS) channel is considered to characterize the upper bound the received SNR.

In this case, we assume that each channel involved is a LOS channel with both a large-scale pathloss and a fixed phase shift. Specifically, the channel from the Tx antenna to the $m$-th REs is denoted by $h_{2,m}=\rho_{2} e^{j\omega_{h2,m}}$, while the channel from the $m$-th REs to the Rx antenna is denoted by $g_{m}=\rho_ge^{j\omega_{g,m}}$.
Then, the optimal phase design at the RIS is obtained from \eqref{eq:Opt_Ref} with $\theta_m = - (\omega_{h2,m} + \omega_{g,m})$. Under these channel realizations, the optimization problem is rewritten as
\begin{subequations}
\begin{align}
\max_{a_m} & \quad \frac{ p_\mathrm{t} \rho^2_2 \rho^2_g (\sum_{m=1}^{M}a_m)^2 }{\rho^2_g\sigma_2^2 \sum_{m=1}^{M}a^2_m+\sigma_1^2}\\
\mathrm{s.t.}&\quad \sum_{m=1}^{M}a_m^2\leq\frac{\upsilon\left(P_\mathrm{RIS}-M (P_\mathrm{c}+P_\mathrm{DC})\right)}{p_\mathrm{t}\rho_2^2+\sigma_2^2}  \label{eq:M_C1}\\
&\quad a_m\leq a_{m,\max},~\forall m. \label{eq:M_C2}
\end{align}
\end{subequations}
According to Cauchy inequality, we have
\begin{align}\label{eq:gamma_opt_Cauchy}
\gamma_{s,a} & \leq \frac{ p_\mathrm{t} \rho^2_2 \rho^2_g M \sum_{m=1}^{M}a^2_m }{\rho^2_g\sigma_2^2 \sum_{m=1}^{M}a^2_m+\sigma_1^2}.
\end{align}
The equality will hold iff the equation $a_1=\dots=a_M$ holds, which means that each RE deployed at the active RIS provides the same amplification gain. This equal amplification gain strategy is straightforward as for the LOS case each RE has the same large-scale pathloss. Therefore, in the following, we consider that each RE is identical with the amplitude gain $a$ and has the same maximum amplitude gain $a_{\max}$. Then, the maximization problem is simplified as
\begin{subequations}
\begin{align}
\max_{a} & \quad \frac{ p_\mathrm{t} \rho^2_2 \rho^2_g M^2a^2 }{\rho^2_g\sigma_2^2 Ma^2+\sigma_1^2}\\
\mathrm{s.t.}&\quad a\leq\sqrt{\frac{\upsilon\left(P_\mathrm{RIS}-M (P_\mathrm{c}+P_\mathrm{DC})\right)}{M(p_\mathrm{t}\rho_2^2+\sigma_2^2)} } \label{eq:MM_C1}\\
&\quad a\leq a_{\max}. \label{eq:MM_C2}
\end{align}
\end{subequations}
Obviously, the objective function is increasing with respect to $a$, and thus the optimal SNR is obtain with $a = \mathrm{min}\left\{\sqrt{\frac{\upsilon\left(P_\mathrm{RIS}-M (P_\mathrm{c}+P_\mathrm{DC})\right)}{M \left( p_\mathrm{t}\rho_2^2+\sigma_2^2\right)}}, a_{\max}\right\}$.

\remark For the proposed LOS channel model, the optimal $\mathbf{\Phi}$ design is to choose $\theta_m = - (\omega_{h2,m} + \omega_{g,m})$ and \begin{equation}\label{eq:M_amp}
a_m = \mathrm{min}\left\{\sqrt{\frac{\upsilon\left(P_\mathrm{RIS}-M (P_\mathrm{c}+P_\mathrm{DC})\right)}{M \left( p_\mathrm{t}\rho_2^2+\sigma_2^2\right)}}, a_{\max}\right\}, \forall m.
\end{equation}

\proposition In LOS case, there exists a trade off between the number of active REs and the amplification power at each element. For the optimal SNR in the active RIS aided system, the optimal number of the active RE $M$ is chosen from  $\lfloor M^{\star} \rfloor$ and $\lceil  M^{\star} \rceil$, where $M^{\star} = \max \left\{M_1,~M_2\right\}$ with $M_1$ and $M_2$ being presented on the top of the next page.

\begin{figure*}[!t]
  \normalsize
  \begin{equation}
  \label{eq:M_1}
  M_1 = \frac{\upsilon P_\mathrm{RIS} \rho^2_g \sigma_2^2 +p_\mathrm{t}\rho^2_2 \sigma_1^2 +\sigma_1^2 \sigma_2^2 - \sqrt{(\upsilon P_\mathrm{RIS} \rho^2_g \sigma_2^2 +p_\mathrm{t}\rho^2_2 \sigma_1^2 +\sigma_1^2 \sigma_2^2)(p_\mathrm{t}\rho^2_2 \sigma_1^2 +\sigma_1^2 \sigma_2^2)}}{\upsilon (P_\mathrm{c}+P_\mathrm{DC})\rho^2_2\rho_g^2}.
  \end{equation}
  \begin{equation}
  \label{eq:M_2}
  M_2 = \frac{\upsilon P_\mathrm{RIS}}{a^2_{\max}(p_\mathrm{t}\rho_2^2+\sigma_2^2)+\upsilon(P_\mathrm{c}+P_\mathrm{DC})}.
  \end{equation}
  \hrulefill
  \vspace*{4pt}
\end{figure*}

\begin{IEEEproof}
First, we first consider the case where the constraint \eqref{eq:MM_C1} is active, and thus we have $M\geq M_2$ and $a = \sqrt{\frac{\upsilon\left(P_\mathrm{RIS}-M (P_\mathrm{c}+P_\mathrm{DC})\right)}{M(p_\mathrm{t}\rho_2^2+\sigma_2^2)}}$. In this case, the optimal SNR is written as
\begin{equation}\label{eq:case1}
  \gamma_{s,a} =\frac{-\upsilon p_\mathrm{t}(P_\mathrm{c}+P_\mathrm{DC})\rho^2_2\rho_g^2 M^2 + \upsilon p_\mathrm{t}P_\mathrm{RIS}\rho_2^2\rho_g^2 M}{-\upsilon (P_\mathrm{c}+P_\mathrm{DC})\rho^2_g \sigma_2^2 M + \upsilon P_\mathrm{RIS} \rho^2_g \sigma_2^2 +p_\mathrm{t}\rho^2_2 \sigma_1^2 +\sigma_1^2 \sigma_2^2}.
\end{equation}
By investigating the first derivative of \eqref{eq:case1}, it is verified that the SNR is increasing when $M\leq M_1$ and decreasing when $M>M_1$. Thus, it reaches the optimal point when $M=M_1$.

Then, we consider the case where the constraint \eqref{eq:MM_C2} is active, and thus we have $M < M_2$ and $a = a_{\max}$. In this case, the optimal SNR is written as
\begin{equation}\label{eq:case2}
  \gamma_{s,a} = \frac{ p_\mathrm{t} \rho^2_2 \rho^2_g M^2a_{\max}^2 }{\rho^2_g\sigma_2^2 Ma_{\max}^2+\sigma_1^2},
\end{equation}
which is obviously increasing with $M$. Thus, it reaches the optimal point when $M = M_2$.

After obtaining the optimal solution for each case, we next investigate the optimal solution to original problem. Notice that, we herein treat the discreet variable $M$ as a continuous one, and therefore the SNR function is also continuous with respective to $M$, which means that the function does not have any abrupt changes in value.

For $M_1\leq M_2$, it is obvious that when $M< M_2$ the constraint \eqref{eq:MM_C2} is active, and thus the SNR is increasing with $M$; when $M\geq M_2$ the constraint \eqref{eq:MM_C1} is active, and thus the SNR is decreasing with $M$. Therefore, the optimal value is obtained with $M^{\star}=M_2$.

For $M_1> M_2$, it is obvious that when $M< M_2$ the constraint \eqref{eq:MM_C2} is active, and thus the SNR is increasing with $M$; when $ M_2\leq M<M_1$, the constraint \eqref{eq:MM_C1} is active, and thus the SNR is increasing with $M$; when $M\leq M_1$ the constraint \eqref{eq:MM_C1} is active, and thus the SNR is decreasing with $M$. Therefore, the optimal value is obtained with $M^{\star}=M_1$.

Overall, we have $M^{\star}=\max\{M_1,M_2\}$. Since $M$ should be an integer, the optimal number of the active REs $M$ is chosen from $\lfloor M^{\star} \rfloor$ and $\lceil  M^{\star} \rceil$. The proof is complete.
\end{IEEEproof}

For the passive RIS, the received SNR increases with the number of REs, and thus the optimal received SNR is written as
\begin{align}\label{eq:Gamma_s_pasive_opt}
  \gamma_{s,p}^{\star} = \frac{p_\mathrm{t} P_\mathrm{RIS}^2\rho_2^2\rho_g^2}{P_\mathrm{c}^2\sigma_1^2}.
\end{align}

The LOS case characterizes an upper bound of the received SNR that the proposed active RIS-aided system can achieve, as it corresponds to the best channel condition and the optimal RIS reflecting coefficient matrix $\mathbf{\Phi}$ is adopted in this case. Although this work just investigates the effects of $M$ on the active RIS with a simplified LOS case, the solution to this simplified problem also has a good performance for more general cases, which has been verified through the numerical results shown in the next section.

\begin{figure}[t]
  \centering\includegraphics[width=.8\columnwidth]{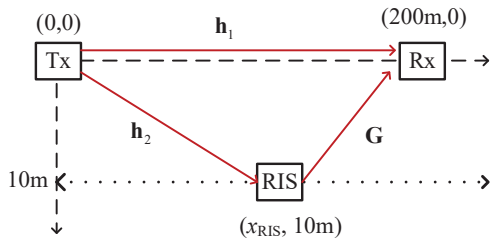} %
  \caption{Simulation setup of the RIS-aided system.}\label{fig:SimModel}
\end{figure}

\section{Simulation Results}\label{sec:Simulation}

In this section, numerical results are provided to investigate the performance of the proposed active RIS SIMO system as illustrated in Fig. \ref{fig:SimModel}. Specifically, we consider a two-dimensional Cartesian coordinate, in which Tx and Rx are located at $(0,0)$ and $(200\mathrm{m},0)$, respectively. Rx is assumed to be located at $(x_\mathrm{RIS},10\mathrm{m})$, which means that the RIS is allowed to move along the horizontal line that is parallel to the $x$-axis. Unless otherwise stated, we set $x_\mathrm{RIS}=180~\mathrm{m}$. Each channel response involved consists of a large-scale fading component and a small-scale fading component. The large-scale fading component is defined by
\begin{equation}\label{eq:LargeScale}
  \rho^{2}_{i}(d_{i}) = \frac{1}{d_i^{\eta_i}\beta_i },
\end{equation}
where $i\in\left\{1, 2, g\right\}$ is the index with $1$, $2$ and $g$ representing the Tx-Rx, Tx-RIS and RIS-Rx channels, respectively, $d_i$ is the distance for the channel $i$, $\beta_i$ is the pathloss at the reference distance (1m) and $\eta_i$ is the pathloss exponent. We assume that $\beta_i = 30~\mathrm{dB}$ for all channels and $\eta_1 = 3.5$, $\eta_2 = 2.8$ and $\eta_g = 2$. Without loss of generality, we assume the Rayleigh fading model to characterize the small-scale fading. All the simulation results related to Rayleigh channel are obtained by averaging over 1000 channel realizations. The power of background noises at the Rx and the active RIS are set to be $\sigma_1^2=\sigma_2^2=-80~\mathrm{dBm}$.

\begin{figure}[t]
    \centering\includegraphics[width=.99\columnwidth]{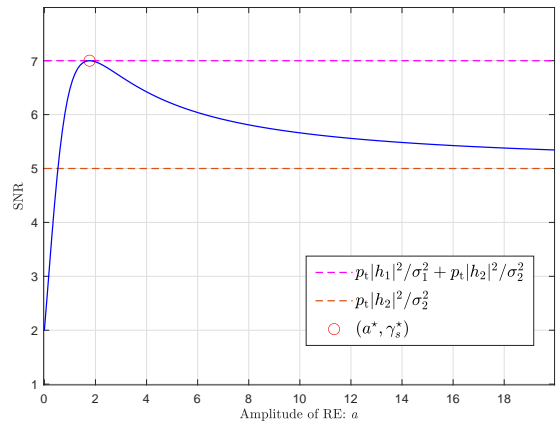} %
    \caption{Received SNR at Rx versus the amplitude of RE: $N=1$, $M=1$ $p_\mathrm{t}/\sigma_1^2 = 10 ~\mathrm{dB}$ and $\sigma_1^2=\sigma_2^2$.} \label{fig:varA}
\end{figure}

First, we investigate the amplification effect on the received SNR in a RIS-aided SISO system where the active RIS is equipped with a single RE. As we are interested in its amplitude in Fig.~\ref{fig:varA}, we assume the phase of the RE is optimized via \eqref{eq:Opt_phase}. In addition, we assume the channels are fixed with the channel gain shown as $|h_1|^2 = 0.2$, $|h_2|^2=0.5$ and $|g|^2=0.8$. It is observed that the received SNR increases dramatically as the amplitude of the RE increases, until it reaches the optimal value $\gamma_s^{\star}=p_\mathrm{t}\left(\frac{|h_1|^2}{\sigma_1^2}+\frac{|h_2|^2}{\sigma_2^2}\right)$. Thus, compared to the case without RIS, the active RIS at most enhances the SNR by $p_\mathrm{t}\frac{|h_2|^2}{\sigma_2^2}$, which can be interpreted as the SNR improvement benefited from the additional channel (with an effective channel gain of  $\frac{|h_2|^2}{\sigma_2^2}$) provided by the active RIS. After the peak point, the received SNR smoothly decreases with the increase in the amplitude and finally converges to $p_\mathrm{t}\frac{|h_2|^2}{\sigma_2^2}$. This is because although with the larger amplitude, the active RIS amplifies the incident signal more, the correlated noise power at the Rx increases more. However, provided that the direct link is totally blocked, increasing the amplitude of the RE always helps to increase the received SNR. The above results demonstrate the importance of optimizing the amplitude of the active RE.

\begin{figure}[t]
  \centering\includegraphics[width=.99\columnwidth]{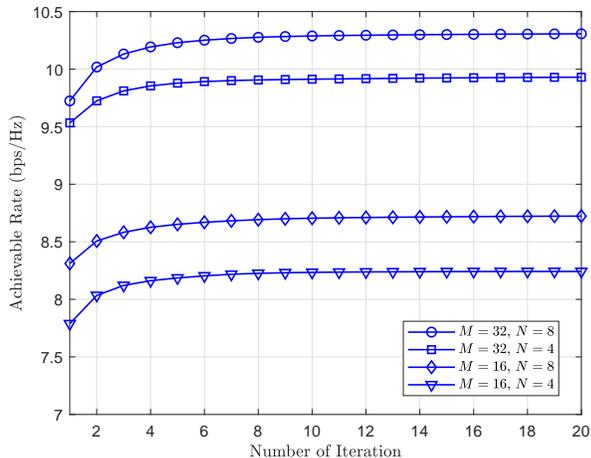} %
  
  \caption{Convergence behaviour of the proposed alternating optimization algorithm: the transmit power $p_\mathrm{t} = 23~\mathrm{dBm}$, the amplification power budget $P_\mathrm{out} = 10~\mathrm{dBm}$, and the amplitude gain $a_{\max}^2 = 40~\mathrm{dB}$.} \label{fig:Converge}
\end{figure}

{Before investigating the performance of the proposed RIS-aided SIMO system, we first show the convergence of the proposed alternating optimization algorithm in Fig.~\ref{fig:Converge} when the transmit power is $p_\mathrm{t} = 23~\mathrm{dBm}$, the amplification power budget is $P_\mathrm{out} = 10~\mathrm{dBm}$ and the maximum amplification gain $a_{\max}^2 = 40~\mathrm{dB}$\footnote{It has been shown in \cite{Amato2018}, the active load can realize $40~\mathrm{dB}$ gain with only $45~\mu\mathrm{W}$ DC power consumption}. The initial point is obtained by setting $\mathbf{\Phi}_0 = a_{\max}\mathbf{I}_{M}$. It is observed that the achievable rate increases quickly with the number of iteration, and finally converges to a value within around 8 iterations for different $(M,N)$ RISs.}

\begin{figure}[t]
  \centering\includegraphics[width=.99\columnwidth]{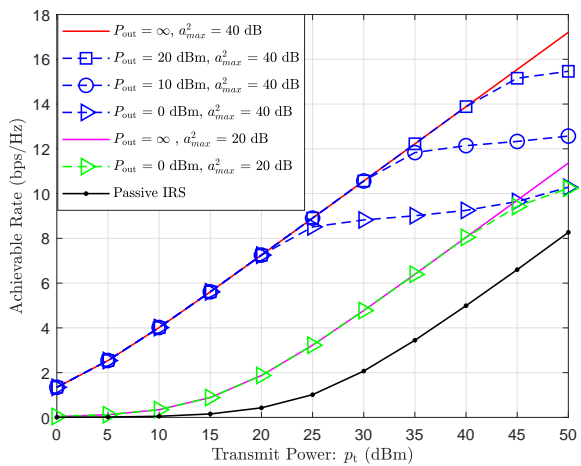} %
  \caption{Achievable rate versus Tx transmit power: $N=4$, $M=16$.} \label{fig:SE}
\end{figure}

In Fig.~\ref{fig:SE}, we plot the achievable rate of the RIS-aided system with the proposed alternating optimization algorithm versus the transmit power at the Tx when the amplification power budget is $P_\mathrm{out} = 20,~10$ and $0~\mathrm{dBm}$, respectively. Generally, the active RIS aided system under various parameter setups outperforms the passive one with the same number of RE, as the RE in the active RIS can amplify the incident signal. For a given $a^2_{\max}=40~\mathrm{dB}$, it is observed that when the transmit power is small, the RIS-aided system with different amplification power budget has almost the same achievable rate, which implies that the amplification power budget constraint is inactive for the weak transmit power. However, as the transmit power increases, the blue dashed curves with different $P_\mathrm{out}$ are distinguished from the other, and the increase of the achievable rate becomes moderate. According to the power budget constraint \eqref{eq:P1C1}, as the transmit power increases, this constraint will be active, resulting in less amplification gain at the active RIS. Thus, the achievable rate gain provided by the active RIS is limited. Meanwhile, if the active RIS has larger amplification power budget, the RIS-aided system has better performance, as it can provide more amplification gain with the same transmit power at the Tx. Therefore, when designing the active RIS-aided system, the RIS has to reconfigure its amplification gain according to the transmit power.

{In addition, for a given $P_\mathrm{out}$, compared with the curves with different amplitude gain $a^2_{\max}$, the curves with the higher amplitude gain have better performance. Especially when the transmit power is small, the rate performance is sensitive to the amplitude gain. This is because when the transmit power is weak, the incident signal at the active RIS is of weak power, and thus the RE can easily amplify the weak incident signal with high amplitude gain but less power consumption. Consequently, when the transmit power is large, the RE has to consume more power to amplify the incident signal, and thus the active RIS-aided system is limited by the amplification power constraint as aforementioned. This is also the reason that the curves with the same $P_\mathrm{out}=0~\mathrm{dBm}$ tends to converge when the transmit power is large.}


\begin{figure}[t]
  \centering\includegraphics[width=.99\columnwidth]{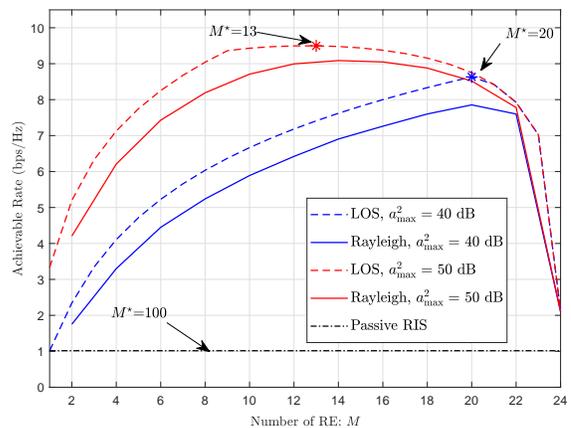} %
  \caption{Achievable rate versus the number of REs $M$: the number of receive antenna $N=1$, the amplifier efficiency $\upsilon=0.8$, the transmit power $p_\mathrm{t}=23~\mathrm{dBm}$, the RIS power budget $P_\mathrm{RIS}=10~\mathrm{dBm}$, the RE circuit power consumption $P_\mathrm{c}=-10~\mathrm{dBm}$, the DC power consumption for each active RE $P_\mathrm{DC} = - 5~\mathrm{dBm}$.} \label{fig:varM}
\end{figure}
Under the same RIS power budget, Fig.~\ref{fig:varM} compares the achievable rate of the RIS-aided system with LOS and Rayleigh channels in terms of the number of RE. In this setup, the passive RIS can employ $100$ REs at most to reflect the incident signal. However, even with 100 REs, the achievable rate with the passive RIS is less than that with the active RIS, as the active RIS can directly amplify the incident signal. On the contrary, the active RIS only takes 13 and 20 REs to reach its optimal performance under different amplitude gains. With the less REs, the active RIS is of smaller surface size and more suitable to the space-limited scenarios. In addition, it is observed that with the increase in $M$, the achievable rate for the active RIS first increases and then decreases after reaching the optimal point, which characterizes the tradeoff between the number of RE and the amplification power. The active RIS can enhance the incident signal by increasing not only the number of REs but also the amplification power. In this power limited case, the active RIS-aided system can benefit more from the increase in $M$ at first, but when $M\geq M^{*}$, it suffers more from the increase in $M$, as more REs reduce the power left for amplification. This result is somehow surprisingly, as it is quite different from the common sense on the passive RIS, that more REs always benefit. It alerts us that $M$ is an important variable that needs to be carefully designed if the RIS has the ability to amplify the incident signal.

Fig.~\ref{fig:varM} also validates the theoretical results proposed in Section \ref{sec:M}. The LOS cases characterize an upper bound of the achievable rate with respect to $M$. It is shown that the optimal $M^{*}$ based on the LOS channel still has a pleasant performance for the proposed alternating optimization algorithm with Rayleigh fading.

\begin{figure}[t]
  \centering\includegraphics[width=.99\columnwidth]{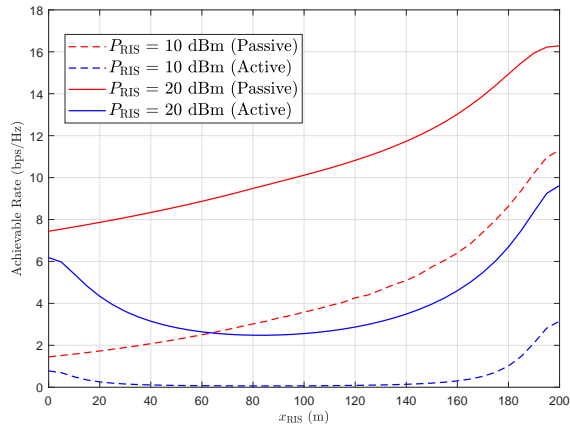} %
  \caption{Achievable rate versus the location of the RIS, where the RIS is located at $(x_\mathrm{RIS},10\mathrm{m})$: the amplifier efficiency $\upsilon=0.8$, the transmit power $p_\mathrm{t}=23~\mathrm{dBm}$, the maximum amplitude gain $a_\mathrm{max}=40~\mathrm{dB}$, the RE circuit power consumption $P_\mathrm{c}=-10~\mathrm{dBm}$, the DC power consumption for each active RE $P_\mathrm{DC} = - 5~\mathrm{dBm}$.} \label{fig:vard}
\end{figure}

In Fig.~\ref{fig:vard}, we compare the achievable rate of the RIS-aided system versus the location of the RIS. These curves are plotted based on the results in Section \ref{sec:M}, with the optimized $M$ in Proposition 2 for the active RIS case. In general, the active RIS has a better performance than the passive RIS with the same RIS power budget, even for the case where the passive RIS is equipped with $1000$ REs when $P_\mathrm{RIS}=20~\mathrm{dBm}$. It is also observed that the passive RIS has the worst performance when the RIS is located close to the middle point between the Tx and Rx (i.e., $x_\mathrm{RIS}=80 \mathrm{m}$), as the double-fading attenuation deteriorates most at this point. However, the active RIS is affected less by the double-fading attenuation, and the achievable rate increases with increasing $x_\mathrm{RIS}$. This is because as the RIS comes closer to the Rx, the signal power of the incident signal at the RIS is weaker, and then, according to \eqref{eq:P1C1}, the active RIS can provide more amplification gain, which compensates for the attenuation caused by the double-fading pathloss. Thus, this interesting result shows that for the active RIS, it is better to deploy it close to the Rx.

\section{Conclusion}\label{sec:Conclusion}
In this paper, we propose a novel RIS-aided SIMO communication system to enhance the spectrum and energy efficiency with fewer REs by leveraging the active RIS via reflecting the incident signal with power amplification. Specifically, the REs in the active RIS exploit active loads to enhance the reflected signal without significantly affecting the low power budget. Based on the proposed system, we have formulated a joint receive and reflect beamforming optimization problem to maximize the received SNR under the RIS power budget constraint. To solve the nonconvex optimization, we have proposed an efficient alternating optimization algorithm which iteratively optimizes the receive beamforming with MMSE criterion and the reflect beamforming with the sequential convex approximation. In addition, we have studied the impact of the number of REs on the received SNR to address the question that how many active REs needs to be deployed. Finally, we have presented numerical results to show that under a practical power consumption model, the proposed active RIS-aided system achieves better performance over the conventional passive RIS assisted system with the same power budget.

\bibliographystyle{IEEEtran}
\bibliography{library.bib}
\end{document}